\documentclass[runningheads]{llncs}

\usepackage[english]{babel}

\usepackage{amsmath}
\usepackage{mathtools}

\let\vec\relax
\DeclareMathAccent{\vec}{\mathord}{letters}{"7E} 
\usepackage{accents}

\usepackage{tikz}
\usetikzlibrary{positioning,shapes,shadows,arrows,backgrounds,automata}
\usepackage{pgf} 

\usepackage{esvect}

\usepackage{paralist}
\usepackage{enumitem}
\usepackage{booktabs}

\usepackage[shadow, textsize=tiny, textwidth=4cm]{todonotes}

\usepackage{amssymb}
\usepackage{stmaryrd} 

\usepackage{cancel}

\newtheorem{privacyrequirement}{\bf Privacy Requirement}

\usepackage{hyperref}

\usepackage{xspace}

\usepackage{graphicx}

\usepackage{mathpartir}

\usepackage{tikz}
\usetikzlibrary{calc}

\usepackage{msc}

\usepackage{tabularx}

\usepackage{multirow}

\usepackage{colortbl}

\usepackage{subcaption}

\usepackage{cite}

\usepackage{wrapfig}

\newcommand{\dataprivacydaynum}{\ensuremath{\mathit{19\hspace{-0.5mm}/\hspace{-0.5mm}04\hspace{-0.5mm}/\hspace{-0.5mm}2024}}\xspace}

\newcommand{\gdprdaynum}{\ensuremath{\mathit{21\hspace{-0.5mm}/\hspace{-0.5mm}12\hspace{-0.5mm}/\hspace{-0.5mm}2024}}\xspace}

\newcommand{\cookiesecure}{\ensuremath{\mathit{cookie{\mkern-2mu}.Secure}}\xspace}

\newcommand{\cookievalue}{\ensuremath{\texttt{c}\xspace}}
\newcommand{\cookieitem}{\ensuremath{\mathit{cookie_{Alice}}\xspace}}
\newcommand{\age}{\ensuremath{\mathit{age}}\xspace}

\newcommand{\varaddress}{\ensuremath{\mathit{address}}\xspace}
\newcommand{\city}{\ensuremath{\mathit{city}}\xspace}

\newcommand{\dt}{\ensuremath{\mathit{t}}\xspace}
\newcommand{\datatype}{\ensuremath{\mathit{t}}\xspace}

\newcommand{\sndr}{\ensuremath{\mathit{sndr}}\xspace}
\newcommand{\rcv}{\ensuremath{\mathit{rcv}}\xspace}
\newcommand{\st}{\ensuremath{\mathit{st}}\xspace}

\newcommand{\cookie}{\ensuremath{\mathit{cookie}}\xspace}

\newcommand{\research}{\ensuremath{\mathit{research}}\xspace}
\newcommand{\advertisement}{\ensuremath{\mathit{advertisement}}\xspace}
\newcommand{\newsletter}{\ensuremath{\mathit{newsletter}}\xspace}

\newcommand{\specialoffers}{\ensuremath{\mathit{special\_offers}}\xspace}
\newcommand{\hoteladvertisement}{\ensuremath{\mathit{hotel\_ads}}\xspace}

\newcommand{\rt}{\ensuremath{\mathit{rt}}\xspace}
\newcommand{\dur}{\ensuremath{\mathit{dur}}\xspace}
\newcommand{\dcr}{\ensuremath{\mathit{dcr}}\xspace}
\newcommand{\TR}{\ensuremath{\mathit{TR}}\xspace}
\newcommand{\tr}{\ensuremath{\mathit{tr}}\xspace}
\newcommand{\pds}{\ensuremath{\mathit{p}_{\ds}}\xspace}
\newcommand{\pdc}{\ensuremath{\mathit{p}_{\dc}}\xspace}
\newcommand{\msgs}{\ensuremath{\mathit{msgs}}\xspace}

\newcommand{\Google}{\ensuremath{\mathit{Google}}\xspace}
\newcommand{\Alphabet}{\ensuremath{\mathit{Alphabet}}\xspace}

\newcommand{\AdsCom}{\ensuremath{\mathit{AdsCom}}\xspace}

\newcommand{\flightsdotcom}{\ensuremath{\mathit{flights.com}}\xspace}
\newcommand{\hotelsdotcom}{\ensuremath{\mathit{hotels.com}}\xspace}

\newcommand{\Alice}{\ensuremath{\mathit{Alice}}\xspace}

\newcommand{\Values}[1]{\ensuremath{\mathcal{V}_{#1}}\xspace}
\newcommand{\DataTypes}{\ensuremath{\mathcal{T}}\xspace}
\newcommand{\DataItems}{\ensuremath{\mathcal{I}}\xspace}
\newcommand{\Purposes}{\ensuremath{\mathcal{P}}\xspace}
\newcommand{\Functions}{\ensuremath{\mathcal{F}}\xspace}

\newcommand{\DataControllers}{\ensuremath{\mathcal{E}}\xspace}

\newcommand{\Entities}{\DataControllers}

\newcommand{\nat}{\ensuremath{\mathbb{N}}\xspace}

\newcommand{\wfc}{\ensuremath{\mathcal{C}}\xspace}

\newcommand{\wfur}{\ensuremath{\mathcal{D{\mkern-0mu}U{\mkern-1mu}R}}\xspace}
\newcommand{\wfcr}{\ensuremath{\mathcal{D{\mkern-0mu}C{\mkern-0mu}R}}\xspace}

\newcommand{\wfpp}[1]{\ensuremath{\mathcal{PP}_{#1}}\xspace} 

\newcommand{\Devices}{\ensuremath{\mathcal{D}}\xspace}
\newcommand{\Events}{\ensuremath{E}\xspace}
\newcommand{\States}{\ensuremath{\mathcal{S}}\xspace}

\newcommand{\PolicySet}[1]{\ensuremath{\policies_{#1}}\xspace} 
\newcommand{\ReceivedDataSet}[1]{\ensuremath{\receiveddata_{#1}}\xspace} 

\newcommand{\request}{\ensuremath{\mathit{request}}\xspace}
\newcommand{\send}{\ensuremath{\mathit{send}}\xspace}
\newcommand{\transfer}{\ensuremath{\mathit{transfer}}\xspace}

\newcommand{\policies}{\ensuremath{\pi}\xspace}

\newcommand{\dataval}{\ensuremath{\nu}\xspace}

\newcommand{\receiveddata}{\ensuremath{\rho}\xspace}

\newcommand{\poentities}{\ensuremath{\leq_{\Entities}}}
\newcommand{\popurposes}{\ensuremath{\leq_{\Purposes}}}
\newcommand{\podata}{\ensuremath{\leq_{\DataTypes}}}

\newcommand{\dursubsumes}{\ensuremath{\preceq_{\wfur}}}
\newcommand{\dcrsubsumes}{\ensuremath{\preceq_{\wfcr}}}

\newcommand{\subsumes}{\ensuremath{\sqsubseteq}}

\newcommand{\eval}[3]{\ensuremath{\texttt{eval}(#1,#2,#3)}\xspace}
\newcommand{\type}{\texttt{type}\xspace}
\newcommand{\activePolicy}{\texttt{activePolicy}\xspace}
\newcommand{\activeTransfer}{\texttt{activeTransfer}\xspace}

\newcommand{\timestamp}{\texttt{time}\xspace}

\newcommand{\owner}{\texttt{owner}\xspace}

\newcommand{\devtoent}{\ensuremath{\texttt{entity}}\xspace}
\newcommand{\comp}{\ensuremath{\underaccent{\sqsupset}{\accentset{\sqsubset}{\scriptscriptstyle =}}}}

\newcommand{\vareval}{\ensuremath{\eta}\xspace}

\newcommand{\true}{\ensuremath{\mathit{tt}}\xspace}
\newcommand{\false}{\ensuremath{\mathit{ff}}\xspace}

\newcommand{\partialfunction}{\ensuremath{\rightharpoonup}}

\newcommand{\abstractsemantics}{\ensuremath{\mathit{AbstSem}\xspace}}
\newcommand{\directrefinement}{\ensuremath{\mathit{DirectRef}\xspace}}
\newcommand{\indirectrefinement}{\ensuremath{\mathit{IndirectRef}\xspace}}

\newcommand{\sendAction}[1]{\ensuremath{\mathit{#1}^{\uparrow}}}
\newcommand{\receiveAction}[1]{\ensuremath{\mathit{#1}^{\downarrow}}}

\newcommand{\PG}{\ensuremath{\mathit{PG}}\xspace}
\newcommand{\Var}{\ensuremath{\mathit{Var}}\xspace}
\newcommand{\Loc}{\ensuremath{\mathit{Loc}}\xspace}
\newcommand{\Act}{\ensuremath{\mathit{Act}}\xspace}
\newcommand{\Effect}{\ensuremath{\mathit{Effect}}\xspace}
\newcommand{\Eval}{\ensuremath{\mathit{Eval}}\xspace}
\newcommand{\Cond}{\ensuremath{\mathit{Cond}}\xspace}
\newcommand{\Sync}{\ensuremath{\mathit{H}}\xspace}

\newcommand{\sync}{\ensuremath{||}\xspace}

\newcommand{\dc}{\text{DC}\xspace}
\newcommand{\dcs}{\text{DCs}\xspace}
\newcommand{\ds}{\text{DS}\xspace}
\newcommand{\dss}{\text{DSs}\xspace}

\newcommand{\eg}{e.g.}
\newcommand{\ie}{i.e.}

\newcommand{\pilot}{{\sc Pilot}\xspace} 
\newcommand{\tlaplus}{\ensuremath{\text{TLA}^+}\xspace} 

\let\temp\phi
\let\phi\varphi
\let\varphi\temp

\begin{document}

\title{Model-Checking the Implementation of Consent}
\author{
Ra{\'u}l Pardo\inst{1} \and
Daniel Le M{\'e}tayer \inst{2}}
\institute{
IT University of Copenhagen, Denmark
(\email{raup@itu.dk})
\and
Univ Lyon, Inria, INSA Lyon, CITI, F-69621 Villeurbanne, France
}

\maketitle

\begin{abstract}
  Privacy policies define the terms under which personal data may be
  collected and processed by data controllers.
  The General Data Protection Regulation (GDPR) imposes requirements
  on these policies that are often difficult to implement.
  Difficulties arise in particular due to the heterogeneity of
  existing systems (\eg, the Internet of Things (IoT), web technology,
  etc.).
  In this paper, we propose a method to refine high level GDPR
  privacy requirements for informed consent into low-level
  computational models.
  The method is aimed at software developers implementing systems that
  require consent management.
  We mechanize our models in \tlaplus\ and use model-checking to prove
  that the low-level computational models implement the high-level
  privacy requirements; \tlaplus has been used by software engineers
  in companies such as Microsoft or Amazon.
  We demonstrate our method in two real world scenarios: an
  implementation of cookie banners and a IoT system communicating via
  Bluetooth low energy.
\end{abstract}
\looseness -1


\section{Introduction}\label{sec:introduction}
The EU General Data Protection Regulation~\cite{GDPR} (GDPR) adopted in May 2016 has been welcomed by many experts as a step forward for privacy protection. 
The GDPR grants new rights to individuals and imposes new obligations on controllers (hereafter, respectively, data subjects, or \dss, and data controllers, or \dcs, following the GDPR terminology)---with significant penalties that could act as a deterrent even for large companies.
However, its actual impact will depend to a large extent on its interpretation and monitoring, but also on the existence of technical tools to implement  its provisions. 
In this respect, the implementation of consent, which is one of the most widely used legal bases for data processing, raises specific challenges. 
For example, Recital 32 of the  GDPR states that  ``consent should be given by a clear affirmative act establishing a freely given, specific, informed and unambiguous indication of the data subject's agreement to the processing of personal data relating to him or her, such as by a written statement, including by electronic means, or an oral statement.'' 
\looseness -1

In practice, internet users generally have to consent on the fly, when they want to use a service, which leads them to accept mechanically the conditions of the provider (e.g.,~\cite{DBLP:conf/chi/JensenP04}). 
Evidently, this does not meet the above GDPR requirements to provide informed consent.
This situation will become even worse with the advent of the internet of things (``IoT''), which has the potential to extend to the ``real world'' the tracking already in place on the internet.
\looseness -1

A common way to tackle this problem is to design privacy languages where users can define their privacy preferences~\cite{DBLP:conf/wpes/MorelP20}.
First, privacy policy languages appeared as a way for users to express and enforce privacy preferences (e.g.~\cite{S4Pbmb10,BDMNpcifa06,MGLpaactavlpp06,DGJKDelshgpl10,PardoLeMetayerDBSec19}).
For example, users may define who can collect their data or whether data can be transferred to 3rd parties.
After the GDPR appeared, some of these languages became partly outdated, as the GDPR introduced new requirements that were not previously foreseen.
This has produced a wave of new languages focused on enforcement and verification of GDPR requirements, e.g., \cite{TokasO20,DBLP:conf/csfw/KaramiBJ22}.
To enforce GDPR requirements, these languages impose certain architectural constraints in the system implementation; for instance, on the way different parties communicate.
However, in practice, not all systems use the same architecture.
There are systems where devices do not support ``direct'' communication with each other, and ``indirect'' means of communication (e.g., via policy repository) are required.
\looseness -1

This raises the following question: \emph{How can we design and verify that implementations of consent management mechanisms ensure GDPR requirements?}
In this paper, we propose a method to verify that different implementations of consent management systems satisfy GDPR requirements.
We build on the \pilot\ privacy policy language~\cite{PardoLeMetayerDBSec19}---whose syntax was designed to capture the information that users and service providers must provide to satisfy GDPR requirements.
Our method uses program graphs (extended state machines) as design language.
Then, we use a mechanization of \pilot\ semantics and program graphs in \tlaplus~\cite{TLA+} to formally verify whether a given implementation refines \pilot semantics.
We show how to prove (via model-checking) that \pilot semantics satisfy requirements for informed consent, and that an implementation refines \pilot semantics (which implies that it satisfies the same informed consent requirements).
The proposed method is aimed at software engineers, which is the reason why we selected \tlaplus\ as verification tool and extended state machines as design language.
\tlaplus\ has been used in industrial applications (e.g.~\cite{DBLP:journals/cacm/NewcombeRZMBD15,DBLP:conf/icse/HackettRK23}), and state machines is a basic model familiar to most software engineers.
\looseness -1

In summary, our contributions are:
\begin{inparaenum}[i)]
\item a formal semantics for \pilot;
\item a framework for software engineers to design implementations of consent management systems using extended state machines known as \emph{program graphs};
\item two case studies modeling real world consent management scenarios: web cookie banners and an IoT system communicating via Bluetooth low energy; and
\item a mechanization in \tlaplus\ of \pilot\ semantics and the program graphs in the case studies. We use the formalization to verify (via the \tlaplus model-checker) that the \pilot\ semantics satisfies privacy requirements for informed consent, and that our case study implementations refine \pilot\ semantics (and, consequently, satisfy the same requirements for informed consent).
\end{inparaenum}
The \tlaplus\ mechanization is available in the accompanying artifact~\cite{pilot-tla-doi} and repository~\cite{formalization-repo}.
\looseness -1


\section{Privacy Policy Language}
\label{sec:language}
\subsubsection{Basic Definitions.}
We start with a set \Devices of \emph{devices} that can store, process and communicate data---\eg, smartphones, laptops or servers.
Let \Entities denote the set of \emph{entities} such as Google or Alphabet and  $\poentities$ the associated partial order---e.g., since Google belongs to Alphabet we have $\Google \poentities
\Alphabet$. 
Entities include \dcs and \dss.
Every device is associated with an entity.
However, entities may have many devices associated with them.
The function $\devtoent : \Devices \to \Entities $ defines the entity
associated with a given device.
Let \DataItems be a set of \emph{data items} corresponding to the pieces of information that devices communicate.
Let \DataTypes be a set of datatypes and  $\podata$ the associated partial order. 
The function $\type : \DataItems \to \DataTypes$ associates a datatype to each data item.
Examples of datatypes are: age, address, city and clinical records.
Since \city is one of the elements that the datatype \varaddress may be composed of, we have $\city \podata \varaddress$.
We use datatype to refer to the semantic meaning of data items.
We use $\Values{}$ to the denote the set of all values of data items, $\Values{} = (\bigcup_{t \in \DataTypes} \Values{t})$ where $\Values{t}$ is the set of values for data items of type $t$.
A special element $\bot \in \Values{}$ denotes the undefined value.
The device where a data item is created (its source) is called the \emph{owner} device of the data item.
The function $\owner \colon \DataItems \to \Devices$ associates an owner device to each data item.
We denote by \Purposes the set of \emph{purposes} and $\popurposes$ the associated partial order.
For instance, if newsletter is considered as a specific type of advertisement, then we have  $\newsletter \popurposes \advertisement$.
\pilot privacy policies are contextual: they may depend on \emph{conditions} regarding information stored in the devices where they are evaluated.
An example of condition is:
\emph{``Only data from adults may be collected.''}
We use a simple logical language to express conditions.
Let \Functions denote a set of functions, and \emph{terms} $t$ be defined as: $t ::= i \mid c \mid f(\vv{t}) $ where $i
\in \DataItems$ is data item, $c \in \Values{}$ is
a constant value, $f \in \Functions$ is a function, and $\vv{t}$ is a list of terms matching the arity of $f$.
The syntax of the logical language is
$
  \phi ::= t_1 \ast t_2 \mid \neg \phi \mid \phi_1 \wedge \phi_2 \mid \true \mid \false
$
where $\ast$ is an arbitrary binary predicate, $t_1,t_2$ are terms; \true and \false represent respectively
true and false.
For instance, the condition above is modeled as $\age \geq 18$.
We denote the set of well-formed conditions as $\wfc$.
\looseness -1

\subsubsection{Abstract Syntax of \pilot Privacy Policies.}
\pilot\ policies express the conditions under which data can be communicated.
We consider two types of data communication: \emph{data collection} and \emph{data transfer}.
Data collection corresponds to \dcs\ collecting \ds\ information.
A transfer is the event of sending previously collected data to third-party \dcs.
\looseness -1

\begin{definition}[\pilot Privacy Policies Syntax]
Given
$\mathit{entity} \in \Entities$,
$\mathit{Purposes} \in 2^\Purposes$,
$\mathit{condition} \in \wfc$,
$\mathit{retention\_time} \in \nat$, and
$\mathit{datatype} \in \DataTypes$,
the syntax of \pilot policies is defined as follows:
$$
\begin{array}{rcl}
  \pilot \, \mathit{Privacy \, Policy}             & ::= & (\mathit{datatype}, \dcr, \TR) \\
  \mathit{Data \, Communication \, Rule} \, (\dcr) & ::= & \langle \mathit{condition}, \mathit{entity}, \dur \rangle \\
  \mathit{Data \, Usage \, Rule} \, (\dur)         & ::= & \langle \mathit{Purposes}, \mathit{retention\_time} \rangle \\
  \mathit{Transfer \, Rules} (\TR)                 & ::= & \{\dcr_1, \dcr_2, \ldots\}\\
\end{array}
$$
\end{definition}
\looseness -1

We use \wfur, \wfcr, \wfpp{} to denote the sets of data usage rules, data communication rules and \pilot privacy policies, respectively.
The set of transfer rules is defined as the set of sets of data communication rules, $\TR \in 2^\wfcr$.
\looseness -1

The purpose of \textit{data usage rules} is to define the operations that may be performed on the data.
$\mathit{Purposes}$ is the set of allowed purposes and $\mathit{retention\_time}$ is the deadline for erasing the data.
For example, consider the data usage rule $\dur_1 = \langle \{\research\}, \dataprivacydaynum \rangle$; tacitly mapping dates to $\mathbb{N}$.
This rule states that data may be used only for the purpose of research until \dataprivacydaynum.
\looseness -1

\textit{Data communication rules} define the conditions under which data must be collected by or communicated to an entity.
The outer layer of data communication rules---i.e., the condition and entity---should be checked by the sender whereas the data usage rule is to be enforced by the receiver.
The first element, $\mathit{condition}$, imposes constraints on the data item and the context (state of the \ds device);
$\mathit{entity}$ indicates the entity allowed to receive the data;
$\dur$ is a data usage rule stating how $\mathit{entity}$ may use the data.
For example,
$ \langle \age > 18, \AdsCom, \dur_1 \rangle $
states that data may be communicated to the entity AdsCom which may use it according to $\dur_1$ (defined above).
It also requires that $\age$ is greater than $18$; $\age$ may be the data item to be sent or contextual information stored in the device.
\looseness -1

\textit{Transfer rules} form a set of data communication rules specifying the entities to which the data may be transferred.
\looseness -1

\dss and \dcs use \pilot \emph{privacy policies} to describe how data may be used, collected and transferred.
The element $\mathit{datatype}$ indicates the type of data the policy applies to, \dcr defines the collection conditions, and \TR the transfer rules.
Several \pilot policies may be necessary to capture the privacy choices for a datatype.
For instance, a \ds may allow only her employer to collect her data when she is at work, and, when being in a museum, only the museum.
To this end, the \ds must define two policies, one for each location.
A \emph{subsumption} relation $p_1 \subsumes p_2$ indicates that $p_1$ is more restrictive than (or subsumes) $p_2$.
Intuitively, a policy is less restrictive than another if it allows for data items to be used for a larger set of purposes, for a longer period of time or sent/transferred to a larger set of entities.
We refer readers to Appendix~\ref{sec:policy-subsumption} for the formal definition of the subsumption relation, as introduced in~\cite{PardoLeMetayerDBSec19}.
\looseness -1

\subsubsection*{Running Example: Cookie Banners.}
Cookie banners are HTML forms that websites use to let users express their choices regarding the use of cookies. 
Typically, they include: the entities allowed to collect the data, the purposes for which they may use it, the retention period and the allowed transfers to third parties.
\looseness -1

\begin{wrapfigure}{r}{0.5\textwidth}
  \centering
  \vspace*{-8mm}
  \includegraphics[width=0.5\textwidth]{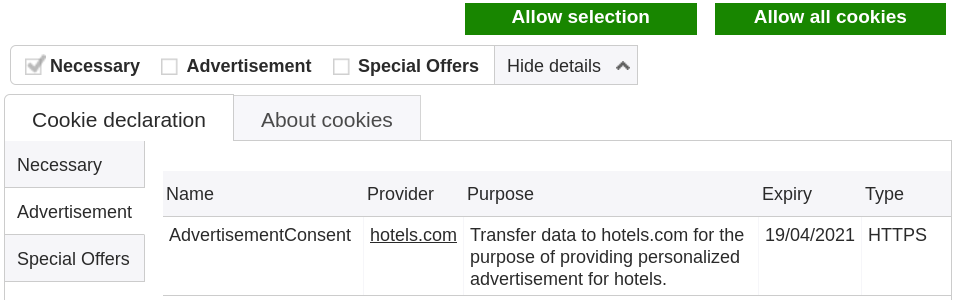}
  \caption{Example Cookie Banner\label{fig:cookie-banner}}
  \vspace*{-8mm}
\end{wrapfigure}

Fig.~\ref{fig:cookie-banner} shows an example where \dss can select the check boxes in the top bar to agree on having the \dc using their data for the specified purposes.
The cookie banner provides detailed information about the types of cookies that the \dc may choose.
Fig.~\ref{fig:cookie-banner} shows information related to advertisement cookies---\dss may click on left tabs to visualize the information related to other cookies.
The main panel indicates the cookie name, provider (\dc), a description of the purpose for data collection, expiration time, and required communication protocol.
Note that sometimes a \dc may set cookies (called third-party cookies) for a different \dc.
\looseness -1

When the GDPR came into force, many \dcs had to adapt their cookie banners to be compliant with the new regulation.
This is all the more important in the view that, according to Recital 42 of the GDPR, \emph{``where processing is based on the data subject's consent, the controller should be able to demonstrate that the \ds has given consent to the processing.''} 
Many websites include unique identifiers in cookies. 
These identifiers may be used for different purposes such as integrity of online sessions, security, statistics or personalized advertisement.
Since users can be uniquely identified using these cookies, according to GDPR, they are \emph{personal data}.
Therefore, \dcs can use them only as necessary for the delivery of their services unless they have obtained a valid consent of the \dss for additional purposes. 
For instance, in the list above, using cookies for integrity of online sessions or communication security does not require explicit consent of the \ds  because it is necessary for the correct functioning of the website.
However,  statistics or advertising are additional purposes for which the consent of the \dss is required. 
In what follows, we show how to use \pilot to encode the information in cookie banners and its advantages as opposed to current cookie banner technology.
%
\looseness -1

Consider the DC, \flightsdotcom, operating a flight search engine website.
The website \flightsdotcom uses cookies to track users and show special offers. 
Furthermore, \flightsdotcom transfers data to a partner company \hotelsdotcom.
The website \hotelsdotcom is a hotel search engine that may use data received from \flightsdotcom to show advertisements, e.g., hotels in users' flight destinations.
Legally speaking, \flightsdotcom must
 obtain the explicit consent of the users to transfer their data to \hotelsdotcom for this purpose.
To this end, \flightsdotcom allows users to choose among four options.
They are shown on left column in Table~\ref{tab:policies}.
\begin{table}[t!]
  \scalebox{0.77}{
    \centering
  \begin{tabular}{ p{2.5mm} p{7.1cm} p{8cm} }
    & \pilot Policy 
    & Textual form for end user GUI
    \\
    \midrule
    1.
    &
    $\bot$
    &
    Collected data may be used only for necessary purposes.
    \\
    \hline
    2.
    &
    $(\cookie,$
    &
    \multirow{2}{8cm}{Data of type \cookie can be collected by \flightsdotcom and used for \specialoffers purposes until $\gdprdaynum$.}
    \\
    &
    $\hspace{0.1cm}\langle \true, \flightsdotcom, \langle \{\specialoffers\}, \gdprdaynum \rangle \rangle,\emptyset)$
    &
    \\
    \hline
    3.
    &
    $(\cookie,$
    &
    \multirow{2}{8cm}{Data of type \cookie can be collected by \flightsdotcom and used for \specialoffers purposes until $\gdprdaynum$. This data may be transferred by \flightsdotcom to \hotelsdotcom which may use it for \hoteladvertisement purpose until $\dataprivacydaynum$.}
    \\
    &
    $\hspace{0.1cm}\langle \true, \flightsdotcom, \langle \{\specialoffers\}, \gdprdaynum \rangle \rangle,$
    &
    \\
    &
    $\hspace{0.2cm}\{\langle \true, \hotelsdotcom, \langle \{\hoteladvertisement\}, \dataprivacydaynum \rangle \rangle\})$
    &
    \\
    &&\\
    \hline
    4.
    &
    $(\cookie, \langle \true, \flightsdotcom, \langle \emptyset, 0 \rangle \rangle,$
    &
    \multirow{2}{8cm}{Data of type \cookie can be collected by \flightsdotcom and transferred to \hotelsdotcom which may use it for \hoteladvertisement purposes until $\dataprivacydaynum$.}
    \\
    &
    $\hspace{0.2cm}\{\langle \true, \hotelsdotcom, \langle \{\hoteladvertisement\}, \dataprivacydaynum \rangle \rangle\})$
    &
    \\
    &&\\
    &&\\
  \end{tabular}}
\caption{\pilot policies for cookie banner.}
\label{tab:policies}
\end{table}

The options correspond to the different configurations shown in Fig.~\ref{fig:cookie-banner}, i.e., selecting:
\begin{inparaenum}[1)]
\item only ``Necessary'';
\item ``Necessary'' and ``Special offers'';
\item ``Necessary'', ``Special offers'' and ``Advertisement'';
\item ``Necessary'' and ``Advertisement''.
\end{inparaenum}
The policy $\bot$ (option 1) corresponds to an \emph{empty policy}, which does not allow \flightsdotcom to use cookies for purposes other than the necessary ones.
The condition (\true) means absence of specific conditions for the cookies to be collected (options 2, 3, 4) or transferred to \hotelsdotcom (options 3 and 4).
The data usage rule $\langle \emptyset, 0 \rangle$ in Option 4 means that \flightsdotcom is not allowed to use cookies for any other purpose than required to deliver its service.
In Option 4, it is only allowed to share information with \hotelsdotcom which can use it to provide hotel advertisements.
This scenario corresponds to the common third-party advertisement business model.
\pilot\ policies can be mapped to natural language for their use in GUIs~\cite{PardoLeMetayerDBSec19}.
We show the natural language version on the right column of Table~\ref{tab:policies}.
\looseness -1

To obtain \ds consent, \flightsdotcom must provide the above \pilot\ policy options the first time that the DS interacts with the website.
A possibility is to attach an XML version of the \pilot policies for each option in the HTML form---see \cite{trustcom:2019} for an encoding of \pilot policies in an IoT system, and not allow DSs to use the website until they make a choice.
Then, the selected policy is sent together with the cookie to inform the DC of the \ds choice.
Alternatively, \dss may define their own \pilot policy before visiting the website.
Consider a \flightsdotcom\ user, Alice, who likes to benefit from special offers.
Alice is a security aware person, and she only agrees to have cookies collected if they are communicated via HTTPS.
Cookies contain the attribute $\mathit{Secure}$, which can be enabled to indicate that they must be transmitted via HTTPS.
To this end, Alice sets the following \pilot policy: 
$(\cookie, \langle \cookiesecure = \true, \flightsdotcom, \langle \{\specialoffers\}, \gdprdaynum \rangle \rangle, $ $\emptyset)$
In practice, Alice would express her policy in its natural language version~\cite{PardoLeMetayerDBSec19}:
\emph{\flightsdotcom may collect data of type \cookie if the attribute \cookiesecure is active and use it for \specialoffers purposes until \gdprdaynum.}
The condition $\cookiesecure = \true$  means that the \pilot policy can be used only if the secure cookie attribute of the cookie used by \flightsdotcom is enabled.
When Alice visits \flightsdotcom, only two options can be selected from the four proposed by \flightsdotcom.
If \flightsdotcom does not use secure cookies, then Alice's policy is not enabled and the first option ($\bot$) must be chosen.
If \flightsdotcom enables secure cookies, then option 2 can also be selected.
By using a language like \pilot, this process can be carried out automatically.

Though implementing cookie banners seems an easy task, it has been shown that many implementations failed to comply with GDPR requirements~\cite{cookiebannerproblems}.
Thus, a remaining question is: How can we assist software engineers in designing and implementing cookie banners that enforce GDPR requirements?
The following sections use this example to describe a method to tackle this problem.
\looseness -1

\section{Abstract Operational semantics}
\label{sec:model}

The \emph{abstract} operational semantics of \pilot defines the conditions under which devices may exchange policies and data.
This semantics does not describe \emph{how} communication or checks are implemented. 
Implementation details are specified using lower level models (cf. \S\ref{sec:refinement}).
We build on the system state introduced in~\cite{PardoLeMetayerDBSec19}.
\looseness -1

Systems are composed of devices that communicate data and use \pilot policies to express \dss and \dcs privacy requirements.
Every device has a set of associated policies.
A policy is associated with a device if it was defined in the device or the device received it.
\ds\ devices have a set of data associated with them.
This models the data stored in the device, e.g., the device MAC address or the user's email.
We keep track of the data collected by \dc devices together with their corresponding \pilot policies.
The system state is formally defined as follows.
\looseness -1

\begin{definition}[System state]\label{def:system-state}
  The system state is a triple $\langle
                                  \dataval,
                                  \policies,
                                  \receiveddata
                                \rangle$
  where:
  \begin{inparaenum}[i)]
  \item
    $\dataval : \Devices \times \DataItems \partialfunction \Values{}$
    is a mapping from the data items of a device to their
    corresponding value in that device.
  \item
    $\policies : \Devices \rightarrow 2^{\Devices \times \wfpp{}}$ is
    a function denoting the \emph{policy base} of a device.
    The policy base contains the policies created by the owner of the
    device and the policies sent by other devices in order to state
    their collection requirements.
    A pair $(d,p)$ means that \pilot policy $p$ belongs to device $d$.
    We write $\PolicySet{d}$ to denote $\policies(d)$.
  \item
    $\receiveddata : \Devices \rightarrow 2^{\Devices \times \DataItems \times
    \wfpp{}}$ returns a set of triples $(s,i,p)$ indicating the data items and
    \pilot policies that a controller has received.
    If $(d',i,p) \in \receiveddata(d)$, we say that device $d$ has received or
    collected data item $i$ from device $d'$ and policy $p$ describes how the
    data item must be used.
    We write $\ReceivedDataSet{d}$ to denote $\receiveddata(d)$.
  \end{inparaenum}
  We denote the universe of all possible states as $\States$.\qed
\end{definition}

In Def.~\ref{def:system-state}, \dataval returns the local value of a data item in the specified device.
When the value of a data item in a device is undefined, \dataval returns $\bot$.
The policy base of a device $d$, $\policies_d$, contains received \pilot policies or policies that have been defined locally.
If $(d,p) \in \policies_d$, the policy $p$ corresponds to a policy that $d$ has
defined in the device itself.
If $(e,p) \in \policies_d$ where $d \not = e$, $p$ is a
policy sent from device $e$.
Policies stored in the policy base are used to compare the privacy policies of two
devices before the data is communicated.
The information that a device has received is recorded in \receiveddata.
Also, \receiveddata contains the \pilot policy describing how data must be used.
The difference between policies in \policies and \receiveddata is that
policies in \policies are used to determine whether data can be
communicated, and policies in \receiveddata are used to describe how a
data item must be used by the receiver.
\looseness -1

\begin{example}\label{ex:system-state}
  Fig.~\ref{fig:state-example} shows a state composed of two devices:
  Alice's web browser, and \flightsdotcom's server.
  The figure depicts the situation after Alice's browser has loaded
  \flightsdotcom and she has chosen an option in the cookie
  banner.
  \looseness -1

  The database in Alice's state ($\dataval_\Alice$) contains a data item of type cookie, \cookieitem, with value \cookievalue.
  The policy base in Alice's browser ($\policies_\Alice$) contains two policies: $(\Alice,p_\Alice)$ representing a policy that Alice defined, and $(\flightsdotcom,$ $p_\flightsdotcom)$ which represents a policy $p_\flightsdotcom$ sent by \flightsdotcom.
  We assume that $p_\Alice$ and $p_\flightsdotcom$ are the policies applying to data items of type cookie.
  \looseness -1

  The state of \flightsdotcom\ contains one more element than Alice's state, namely, the set of received data ($\receiveddata_\flightsdotcom$).
  This set contains the item \cookieitem\ collected from Alice, and the policy $p_\flightsdotcom$ that describing how to handle the data.
  Note that $p_\flightsdotcom$ was the \pilot policy originally defined by \flightsdotcom.
  To satisfy Alice's privacy preferences, $p_\flightsdotcom$ must be more restrictive than Alice's policy $p_\Alice$, which is denoted by $p_\flightsdotcom \subsumes p_\Alice$.
  This condition can easily be enforced by comparing the policies before data is collected.
  The first element in $(\Alice, $ $\cookieitem,$ $p_\flightsdotcom)$ indicates that the data comes from Alice's device.
  Finally, \flightsdotcom's policy base has one policy: its own policy $p_\flightsdotcom$, which was communicated to Alice for data collection.
  \looseness -1
\end{example}


\tikzstyle{abstract}=[rectangle, draw=black, rounded corners, fill=white,
                      text centered, anchor=north, text=black, text width=4.5cm]

\begin{figure}[!t]
\centering
\scalebox{0.65}{
\begin{tikzpicture}
     \node (car)
        {
          \includegraphics[width=2cm]{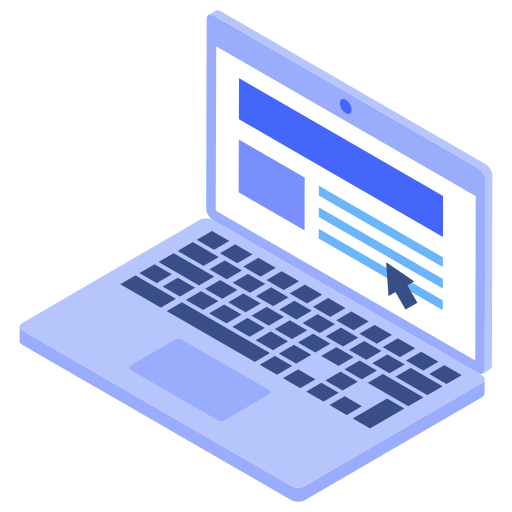}
        };

    \node (piSubject) [abstract,
                       rectangle split,
                       rectangle split parts=3,
                       left=of car,
                       yshift=5mm,
                       xshift=5mm]
        {
          $\boldsymbol{\policies_\Alice}$
          \nodepart{two}$(\Alice,p_\Alice)$
          \nodepart{three}$(\flightsdotcom,p_\flightsdotcom)$
        };
    \node (datavalSubject) [abstract,
                             rectangle split,
                             rectangle split parts=2,
                             below=of piSubject,
                             yshift=8mm
                            ]
        {
          $\boldsymbol{\dataval_\Alice}$            
          \nodepart{two}$(\cookieitem,\cookievalue)$
        };
    \node (aux1)  [below=of piSubject,
                   yshift=1cm]{};
    \begin{pgfonlayer}{background}
      \filldraw [line width=4mm,join=round,black!10]
      (piSubject.north  -| piSubject.east)  rectangle (datavalSubject.south  -| datavalSubject.west);
    \end{pgfonlayer}

    \node (camera-gate) [right=of car,
                         xshift = 2cm,
                         yshift = 0cm]
        {
          \includegraphics[width=3cm]{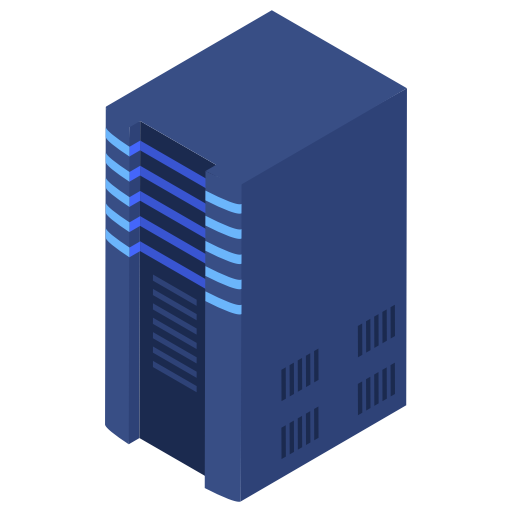}
        };

    \node (datavalController) [abstract,
                               rectangle split,
                               rectangle split parts=2,
                               right=of camera-gate,
                               yshift=0mm,
                               xshift=-10mm]
        {
            $\boldsymbol{\dataval_\flightsdotcom}$
            \nodepart{second}$(\cookieitem,\cookievalue)$
        };

    \node (piController) [abstract,
                          rectangle split,
                          rectangle split parts=2,
                          above = of datavalController,
                          yshift = -8mm]
        {
            $\boldsymbol{\policies_\flightsdotcom}$
            \nodepart{two}$(\flightsdotcom,p_\flightsdotcom)$
        };

    \node (rhoController) [abstract,
                           rectangle split,
                           rectangle split parts=2,
                           below=of datavalController,
                           yshift=8mm]
        {
            $\boldsymbol{\receiveddata_\flightsdotcom}$
            \nodepart{second}$(\Alice,\cookieitem,p_\flightsdotcom)$
          };
     \draw[<-, thick, >=stealth] (car) to [bend right] (camera-gate);
     \draw[<-, thick, >=stealth] (camera-gate) to [bend right] (car);
    \begin{pgfonlayer}{background}
      \filldraw [line width=4mm,join=round,black!10]
      (piController.north  -| piController.east) rectangle (rhoController.south  -| rhoController.west);
    \end{pgfonlayer}

\end{tikzpicture}
}
\caption{Cookie Banner System State Example}
\label{fig:state-example}
\vspace*{-2mm}
\end{figure}
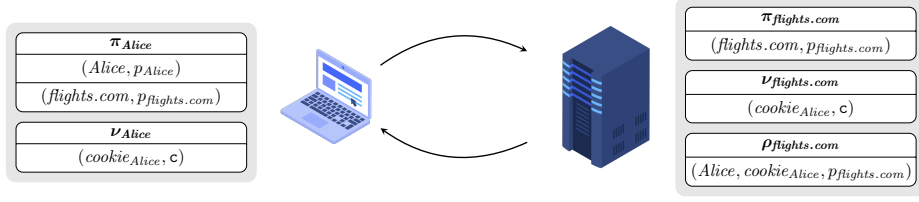


\subsubsection{System Events.}\label{subsec:semantics}

We define the abstract semantics of the events in~\cite{PardoLeMetayerDBSec19} focused on exchange of data items and \pilot policies.
The set of events, denoted as $\Events$, includes: $\request$, $\send$, and $\transfer$.
\emph{Traces} are sequences of state-event pairs $(s_1,e_1),$ $(s_2,e_2),$ $(s_3,e_3), $ $\ldots$ where $s_i \in \States$ is a system state and $e_i \in \Events$ is an event.
A state $s_{i+1}$ is the result of executing $e_i$ on state $s_i$.
We write $s \xrightarrow{e} s'$ to denote that $s'$ is reached by executing $e$
on state $s$.
Every event is tagged with a timestamp indicating when the event occurs.
We use a function $\timestamp : \Events \to \nat$ to assign the timestamp---represented as a natural number.
\looseness -1

Table~\ref{tab:os-com-events} shows the definition of the events using small step operational semantics rules.
The conditions of the rules are written as premises and the state updates are written on the right hand side.
The functions \activePolicy\ and \activeTransfer\ determine whether a policy is active upon sending/transferring a data item.
Intuitively, a policy is active if it applies to the data item to be sent, to the receiving entity, the retention time has not yet been reached, and its condition holds.
The same must hold for active transfers (but for the transfer rule), and also the retention time for the sender (i.e., the retention time in the policy) must not have been reached.
%
%
We refer readers to Appendix~\ref{sec:active-policy-transfer} for the formal definition of these functions, as introduced in~\cite{PardoLeMetayerDBSec19}.
We use $p.t$, $p.\TR$ and $p.\dcr.e$ to refer to the datatype associated to $p$, its set of transfer rules and the entity specified in the communication rule, respectively.
\looseness -1

\begin{table}[t!]  
  \scalebox{0.83}{
  \begin{tabular}{c}
    \multicolumn{1}{l}{\sc Request}
    \\
    $
      \inferrule*[Left=R1,
                  Right=$\mathit{with }$ ${\small \PolicySet{\rcv}' = \PolicySet{\rcv} \cup \{(\sndr,p)\}}$]
                 {
                  (\sndr, p) \in \PolicySet{\sndr}
                  \\
                  \forall (\sndr,q) \in \PolicySet{\rcv} \cdot p \; \cancel{\comp}\; q
                 }
                 {
                  \langle \dataval, \policies, \receiveddata \rangle
                  \xrightarrow{\request(\sndr,\rcv,\datatype, p = (t,\_,\_))}
                  \langle \dataval, \policies', \receiveddata \rangle
                 }
    $
    \\
    \\
    $
      \inferrule*[Left=R2,
                  Right=$\mathit{with }$ ${\small \PolicySet{\rcv}' = (\PolicySet{\rcv} \setminus q) \cup \{(\sndr,p)\}}$]
                 {
                  (\sndr, p) \in \PolicySet{\sndr}
                  \\
                  (\sndr,q) \in \PolicySet{\rcv}
                  \\
                  p \; \comp \; q
                 }
                 {
                  \langle \dataval, \policies, \receiveddata \rangle
                  \xrightarrow{\request(\sndr,\rcv,\datatype, p = (t,\_,\_))}
                  \langle \dataval, \policies', \receiveddata \rangle
                 }
    $
    \\
    \\
    \multicolumn{1}{l}{\sc Send}
    \\
    $
    \inferrule*[Right=$\mathit{with} \hspace*{-1mm}$
                {\small
                   $\begin{cases}
                    \ReceivedDataSet{\rcv}' = \ReceivedDataSet{\rcv} \cup \{(\sndr,i,p_{\rcv})\}
                    \\
                    \dataval'(\rcv,i) = \dataval(\sndr,i)
                  \end{cases}$
                }
                ]
                {
                 (\rcv,p_{\rcv}) \in \PolicySet{\sndr}
                 \\
                 \activePolicy(p_{\rcv},\send(\sndr,\rcv,i),\st)
                 \\\\
                 (sndr,p_{sndr}) \in \PolicySet{\sndr}
                 \\
                 \activePolicy(p_{\sndr},\send(\sndr,\rcv,i),\st)
                 \\\\
                 p_{\rcv} \subsumes p_{\sndr}
                 \\
                 \dataval(\sndr,i) \not = \bot
                }
                {
                 st = \langle \dataval, \policies, \receiveddata \rangle
                 \xrightarrow{\send(\sndr,\rcv,i)}
                 \langle \dataval', \policies, \receiveddata' \rangle
                }
    $
    \\
    \\
    \multicolumn{1}{l}{\sc Transfer}
    \\
    $
     \inferrule*[Right=$\mathit{with} \hspace*{-1mm}$
                {\small
                   $\begin{cases}
                     \ReceivedDataSet{\rcv}' = \ReceivedDataSet{\rcv} \cup \{(\sndr,i,p_{\rcv})\}
                     \\
                     \dataval'(\rcv,i) = \dataval(\sndr,i)
                  \end{cases}$
                }
                ]
                {
                 (\_,i,p) \in \ReceivedDataSet{\sndr}
                 \\
                 \tr \in p.\TR
                 \\\\
                 \activeTransfer(\tr,p,\transfer(\sndr,\rcv,i),\st)
                 \\\\
                 (\rcv,p_\rcv) \in \PolicySet{\sndr}
                 \\
                 \activePolicy(p_\rcv,\transfer(\sndr,\rcv,i),\st)
                 \\\\
                 p_\rcv \subsumes (p.t, \tr, p.\TR)
                 \\
                 \dataval(\sndr,i) \not = \bot
                }
                {
                 \st = \langle \dataval, \policies, \receiveddata \rangle
                 \xrightarrow{\transfer(\sndr,\rcv,i)}
                 \langle \dataval', \policies, \receiveddata' \rangle
                }
    $
    \\
    \\
  \end{tabular}
  }
  \caption{Small Step Operational Semantics of Events.}
  \label{tab:os-com-events}
\end{table}

The event \textit{$\request(\sndr,\rcv,\datatype,p)$} models a data request from a \dc to a \ds or another \dc.
Thus, \sndr is always a \dc device, and \rcv may be a \dc or \ds device.
A request includes the type $\datatype$ of the requested data  and a \pilot policy $p$.
The policy is required to refer to the datatype that is requested, \ie, $p = (\datatype,\_,\_)$.
After executing \request, the pair $(\sndr, p)$ is added to $\PolicySet{\rcv}$ (R1).
If \rcv\ already received a comparable policy from \sndr ($p \, \comp \, q \triangleq p \subsumes q \vee q \subsumes p$), then $q$ is replaced with $p$ (R2); as new \dcs\ policies may be more or less restrictive.
Thus, \rcv is informed about the updated terms for data handling.

The event \textit{$\send(\sndr,\rcv,i)$} models collection of data item $i$ from \ds $\sndr$ by \dc $\rcv$.
Devices can only send $i$ if defined in the state, $\dataval(\sndr,i) \not = \bot$.
To execute \send, it must hold that $\PolicySet{\sndr}$ contains:
\begin{inparaenum}[i)]
\item an active policy defined by \sndr, $p_\sndr$, indicating how \sndr allows \dcs to use her data, and
\item an active policy sent by \rcv, $p_\rcv$, indicating how it commits to use the data.
\end{inparaenum}
Data is sent if $p_\rcv$ is more restrictive than $p_\sndr$ (i.e., $p_\rcv \subsumes p_\sndr$).
We record the exchange in $\ReceivedDataSet{\rcv}$ indicating: the sender, the data item and  \rcv's \pilot policy, $(\sndr,i,p_\rcv)$.
We update \rcv's database with the value of $i$ in \sndr's
state, $\dataval(\rcv,i) = \dataval(\sndr,i)$.

The event \textit{$\transfer(\sndr,\rcv,i)$} models a \dc\ (\sndr) transferring a data item $i$ to another \dc\ (\rcv)---if defined in the state, $\dataval(\sndr,i) \not = \bot$.
First, it must hold that $\PolicySet{\sndr}$ contains an active policy from \rcv, $p_\rcv$.
Let $p$ denote the policy sent by the data owner along with $i$.
There must exist an active transfer rule (\tr) in the set of transfers rules of $p$.
As before, the policy sent by $\rcv$ must be more restrictive than those sent by data owners, i.e., $p_\rcv \subsumes p_\tr$ where $p_\tr$ is a policy with the active transfer $\tr$ in the place of the data communication rule and with the same set of transfers as $p$.
Data items can be transferred more than once to the entities in the set of transfers as long as the retention time has not been reached.
This is not an issue in terms of privacy as data items are constant values.
We update $\ReceivedDataSet{\rcv}$ with the sender, the data item and \rcv's \pilot policy, $(\sndr,i,p_{\rcv})$.
Note that the owner of the data item is not \sndr since transfers always correspond to exchanges of previously collected data, $\owner(i) \not = \sndr$.
The database of \rcv is updated with the current value of $i$ in $\dataval_\sndr$.
\looseness -1



\section{Refinement into Operational Models}
\label{sec:refinement}
A \emph{refinement} is an operational model that can be directly implemented---i.e., it can easily be translated to source code.
Refinements provide information about the \emph{communications} between devices and the \emph{locality of the computations}; these details are omitted in the abstract semantics.
A refinement must include operational models for each device in the system.
We model refinements using a symbolic representation of labeled transition systems called \emph{program graphs} (PGs)~\cite{principles_of_model_checking}.
These models can be seen as extended state machines---and should be familiar to most software engineers, as they are part of their basic education.
For each PG, we consider a (finite) set of variables \Var.
Each variable $x \in \Var$ belongs to a domain $D_x$, we use
$D = \bigcup_{x \in \Var} D_x$ to denote the set of all domains.
Functions $\vareval \colon \Var \to D$ map variables to a value in
their corresponding domain, and we use $\Eval(\Var)$ to denote the set
of all $\vareval$ functions.
We use $\Cond(\Var)$ to denote a set of conditions on variables.
Program graphs are defined as follows:
\looseness -1

\begin{definition}[Program Graph~\cite{principles_of_model_checking}]
  A \emph{program graph} \PG over a set of variables \Var is a tuple
  $(\Loc,\Act, \Effect,\to,\Loc_0,g_0)$ where
  $\Loc$ is a set of locations and $\Act$  is a set of actions;
  $\Effect \hspace*{-0.3mm} \colon \hspace*{-0.3mm} \Act \hspace*{-0.3mm} \times \hspace*{-0.3mm} \Eval(\Var) \hspace*{-0.3mm} \to \hspace*{-0.3mm} \Eval(\Var)$ is an
    effect function;
  $\to_{\PG} \subseteq \hspace*{-0.3mm} \Loc \hspace*{-0.3mm} \times \hspace*{-0.3mm} \Act \hspace*{-0.3mm} \times \hspace*{-0.3mm} \Cond(\Var) \hspace*{-0.3mm} \times \hspace*{-0.3mm} \Loc$ is a conditional transition relation;
  $\Loc_0 \hspace*{-0.7mm} \subseteq \hspace*{-0.7mm} \Loc$ is a set of initial locations; and
  $g_0 \hspace*{-0.7mm} \in \hspace*{-0.7mm} \Cond(\Var)$ is the initial condition.
  \looseness -1
\end{definition}
When no confusion may arise, we omit the subscript in $\to_{\PG}$.
By convention the shape of the actions in $\Act$ is $a(\vec{x})$
where $\vec{x} = x_1, x_2, \ldots$ is a finite vector of variables,
called the \emph{parameters} of the action.
We write $q \xrightarrow{a(\vec{x}) / \phi} q'$ to denote that there exists a
transition from $q$ to $q'$ in the PG labeled by $a(\vec{x}) / \phi$, i.e.,
$(q,a(\vec{x}),\phi,q') \in\ \to_{\PG}$.
For simplicity, when $\phi = \top$ we write
$q \xrightarrow{a(\vec{x})} q'$.

In the following we define the synchronous composition of PGs.
To this end, we use \emph{synchronization actions}.
For any synchronization action $a(\vec{x})$, there are corresponding sending ($\sendAction{a}(\vec{x})$), and receiving ($\receiveAction{a}(\vec{x})$) actions.
Two devices communicate via $a(\vec{x})$ by each of them executing one of the corresponding sending and receiving action.
We use $\vareval[\,\vec{x}_{\receiveAction{a}} \mapsto \vareval(\vec{x}_{\sendAction{a}})\,]$ to denote that the variables in the receiving action are bind with the values of the sending action in \vareval.
We use synchronization actions to formally define the synchronous composition of PGs.
\looseness -1

\begin{definition}[Synchronous Composition]
  Given two PGs $\PG_1$ and $\PG_2$, such that
  $\PG_i = (\Loc_i, \Act_i, \Effect_i, \to_i, \Loc_{0_{i}}, g_{0_{i}})$,
  the \emph{synchronous composition} over the actions 
  $\Sync = \{ a(\vec{x}) \mid \sendAction{a}(\vec{x}) \in \Act_j \wedge \receiveAction{a}(\vec{x}) \in \Act_k 
           \text{ for } j,k \in \{1,2\} \text{ and } j \not= k \}$
  is defined as,
  $\PG_1 \sync_{\Sync} \PG_2 \triangleq (\Loc_1 \times \Loc_2,
                                       \Act_1 \cup \Act_2 \cup H, 
                                       \Effect,
                                       \to,
                                       \Loc_{0_{1}} \times \Loc_{0_{2}},
                                       g_{0_{1}} \wedge g_{0_{2}})$
  and the transition relation $\to$ is defined as
  \[\scriptsize
    \inferrule*[right={\scriptsize $ \mathit{with } ~ i \in \{1,2\}$}]
    { q_i \xrightarrow{a(\vec{x})/\phi_i}_i q_i' \\ a(\vec{x}) \notin \Sync}
    { (q_1,q_2) \xrightarrow{a(\vec{x})/\phi_1} (q'_1,q_2)}
    \hspace{.2cm}
    \inferrule*
    { q_1 \xrightarrow{\sendAction{a}(\vec{x})/\phi_1}_1 q_1' \\
      q_2 \xrightarrow{\receiveAction{a}(\vec{x})/\phi_2}_2 q_2' \\
      a(\vec{x}) \in \Sync
    }
    { (q_1,q_2) \xrightarrow{a(\vec{x})/\phi_1\wedge\phi_2} (q_1',q_2')}
  \]
  and, $\Effect(a(\vec{x}),\vareval) = $
  $\scriptsize
  \begin{cases}
    \Effect_i(a(\vec{x}),\vareval) & \text{if } a(\vec{x}) \notin \Sync \text{ and } 
                                     a(\vec{x}) \in \Act_i
                                     \text{ with } i \in \{1,2\} \\
    \vareval[\,\vec{x}_{\receiveAction{a}} \mapsto \vareval(\vec{x}_{\sendAction{a}})\,] & \text{if } a(\vec{x}) \in \Sync
  \end{cases}
  $
  \label{def:parallelcomposition}  
\end{definition}

In what follows, we use PGs to model implementations of different system architectures.
We model devices that can communicate directly and indirectly.
We consider, w.l.o.g., policies and transfers that are always active; note that this check can always be added as a prerequisite before sending data.
We prove that the implementations refine the abstract semantics of \pilot in \S\ref{sec:verification}.
\looseness -1

\subsubsection{Direct Communication: Cookie Banners.}
\label{subsec:direct-communication}
\tikzset{every loop/.style={looseness=10}}

We return to our cookie banner example.
We define two PGs for the two devices in the example: Alice's laptop
(DS device) and the server hosting \flightsdotcom (DC device).
These PGs precisely describe the program that Alice and the server
must execute in their devices, and the communication between them.
We model a system where the \ds defines a policy \textit{a priori} and data collection is allowed if the \dc policy complies with the \ds policy.
Note that this differs from the usual interaction where the DS is
presented with several DC policies, and she must select one
(cf. Fig.~\ref{fig:cookie-banner}).
\looseness -1

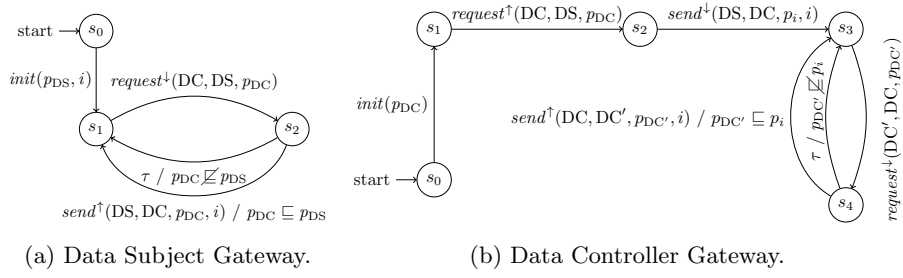
\begin{figure*}[t!]
    \hspace{0mm}
    \begin{subfigure}[b]{0.37\textwidth}
        \centering
        \resizebox{\linewidth}{!}{
            \begin{tikzpicture}[every text node part/.style={align=center},on grid, auto]
        \node (s_0) [initial, circle, draw]%
                    {$s_0$};
        \node (s_1) [circle, draw]%
                    [below = 2cm of s_0]%
                    {$s_1$};
        \node (s_2) [circle, draw]
                    [right = 4cm of s_1]
                    {$s_2$};
        \path[->] [bend left =  0] (s_0) edge node [xshift=-19mm] {$\mathit{init(\pds,i)}$} (s_1);
        \path[->] [bend left = 30] (s_1) edge node {$\receiveAction{\request}(\dc,\ds,\pdc)$} (s_2);
        \path[->] [bend left = 30] (s_2) edge node {$\mathit{\tau~/~p_{\dc} \, \cancel{\subsumes} \, p_{\ds}}$} (s_1);
        \path[->] [bend left = 70] (s_2) edge node {$\mathit{\sendAction{\send}(\ds,\dc,p_{\dc},i)~/~p_{\dc} \subsumes p_{\ds}}$} (s_1);
      \end{tikzpicture}
        }
        \caption{Data Subject Gateway.}
        \label{fig:dsg-direct}
    \end{subfigure}
    \hfill
    \begin{subfigure}[b]{0.62\textwidth}
    \centering
        \resizebox{\linewidth}{!}{
            \begin{tikzpicture}[every text node part/.style={align=center},on grid, auto]
        \node (s_0) [initial, initial where=left, circle, draw]%
                    {$s_0$};
        \node (s_1) [circle, draw]%
                    [above = 3cm of s_0]
                    {$s_1$};
        \node (s_2) [circle, draw]%
                    [right = 4.1cm of s_1]
                    {$s_2$};
        \node (s_3) [circle, draw]%
                    [right = 4.1cm of s_2]%
                    {$s_3$};
        \node (s_4) [circle, draw]%
                    [below = 3.5cm of s_3]%
                    {$s_4$};
        \path[->] (s_0) edge node {$\mathit{\mathit{init}(\pdc)}$} (s_1);
        \path[->] (s_1) edge node [yshift=0mm,xshift=0mm,rotate=0] {$\sendAction{request}(\dc,\ds,\pdc)$} (s_2);
        \path[->] (s_2) edge node [yshift=0mm,xshift=0mm,rotate=0] {$\receiveAction{send}(\ds,\dc,p_{i},i)$} (s_3);
        \path[->] [bend left = 20] (s_3) edge node [rotate=90, xshift=-2.2cm, yshift=-5mm] {$\receiveAction{request}(\dc',\dc,p_{\dc'})$} (s_4);
        \path[->] [bend left = 20] (s_4) edge node [rotate=90,yshift=2mm,xshift=12.5mm] {$\mathit{\tau~/~p_{\dc'} \, \cancel{\subsumes} \, p_{i}}$} (s_3);
        \path[->] [bend left = 60] (s_4) edge node {$\sendAction{send}(\dc,\dc',p_{\dc'},i)~/~p_{\dc'} \subsumes p_{i}$} (s_3);
      \end{tikzpicture}
        }
        \caption{Data Controller Gateway.}
        \label{fig:dcg-direct}
    \end{subfigure}
\caption{Direct Communication Program Graphs.} 
\label{fig:direct_pgs}
\end{figure*}

\textit{Data Subject Gateway (DS program)}.
This program allows data collection only if the DC policy respects (subsumes) Alice's policy.
Fig.~\ref{fig:dsg-direct} shows the data subject gateway PG.
This PG runs on Alice's browser (\ds\ device).
It contains the following variables:
$p_{\ds}$ corresponding to $\policies(\ds,\ds)$ in the abstract
semantics, $p_{\dc}$ corresponding to $\policies(\ds,dc)$, and the
current state $\sigma$ of the PG (not in abstract semantics).
We define \Effect\ as:
$\Effect(\mathit{init}(i,\pds),\vareval) = \vareval[\pds \mapsto \texttt{policy\_init()}, i   \mapsto \texttt{data\_init()}]
$
and
$\Effect(\tau,\vareval) = \vareval.$
The functions \texttt{policy\_init()} and \texttt{data\_init()} model
machine-level functions that retrieve: the policy from the user (\eg,
via a browser extension or settings form), and the data item (\eg,
when entered by the user or automatically retrieved by the browser),
respectively.
For each data item, an instance of the data subject gateway is executed.
Internal actions $\tau$ model transitions with no effect on PG variables.
The synchronization events \request\ and \send\ are defined as in Def.~\ref{def:parallelcomposition}; they synchronize with the corresponding events in the DC program described below.
\looseness -1

Intuitively, the PG in Fig.~\ref{fig:dsg-direct} starts with an initialization action, $\mathit{init}(\pds,i)$, where Alice's inputs her policy (\pds) and the data item ($i$).
Then, the data subject gateway waits for data collection requests,
$\receiveAction{\request}(\dc,\ds,\pdc)$.
A request includes the sender \dc, the receiver \ds, and the policy of
the \dc.
Upon receiving a request, the data subject gateway checks whether the
\dc policy complies with that of the \ds, \pdc \subsumes \pds.
If so, it sends the item ($i$) along with the policy (\pdc) that
must be enforced when processing $i$,
$\mathit{\sendAction{\send}(\ds,\dc,p_{\dc},i)}$.
Otherwise, data collection does not take place, and the PG goes back
to wait for further requests.
\looseness -1

\textit{Data Controller Gateway (\dc program)}.
This program can request data from \dss, process
it according to its privacy policy, and transfer it (if requested by
another \dc and allowed by the policy).
Fig.~\ref{fig:dcg-direct} shows the controller gateway PG.
This PG runs on the server hosting \flightsdotcom.
It consist of the following variables: $p_{i}$ and $i$ correspond to
the policy and data item sent by the \ds:
$ (i,\ds,p_i) \in \receiveddata(\dc)$ in the abstract semantics,
$p_{\dc} = \policies(\dc,\dc)$, $p_{\dc'} = \policies(\dc,\dc')$, and
current state $\sigma$ of the PG (not in abstract semantics).
The \Effect\ function is defined as:
$
\Effect(\mathit{init}(\pdc),\vareval)=\vareval[\pdc \mapsto \texttt{policy\_init()}] \text{ and } \Effect(\tau,\vareval) = \vareval.
$
In this case, \texttt{policy\_init()} retrieves the \dc policy.
\looseness -1

The PG in Fig.~\ref{fig:dcg-direct} starts by initializing the \dc
policy, \pdc.
Then, it sends a request to collect data of the type specified in
\pdc.
This occurs, for instance, when a client visits \flightsdotcom.
At this point, the data controller gateway waits to receive the data
from the \ds.
Upon receiving the data, it can transfer the data to another data
controller $\dc'$.
This corresponds to the edges between $s_3$ and $s_4$.
This interaction is identical to that of the data subject gateway
(cf. Fig.~\ref{fig:dsg-direct}).
The only difference is that the policy used to determine the
compliance with the \ds preferences is $p_i$, \ie, the policy sent by
the \ds during data collection.
Note that this ensures that \ds preferences are enforced even if data
items are transferred---this property is formally proved in \S\ref{sec:verification}.
\looseness -1

This implementation is applicable to systems with support for direct communication.
For example, a system where users upload data from a smartwatch to fitness social networks such as Strava or Fitbit.
However, this implementation cannot be applied to systems that automatically broadcast data---\eg, Bluetooth protocols---where receivers can record \ds data as soon as \ds devices enter the collection range.
Below we describe an alternative refinement for these systems.
\looseness -1

\subsubsection{Indirect Communication: Bluetooth Low Energy.}
\label{subsec:indirect-communication}
Many wireless communication protocols, such as \emph{Bluetooth low energy} (BLE), require broadcasting identifying information such as MAC addresses to operate.
Unfortunately, this information is becoming a common way to track \dss in public places~\cite{trustcom:2019, OOSTERLINCK201755, DBLP:journals/sensors/DanisC17, DBLP:journals/popets/BeckerLS19}.
This type of data collection is especially dangerous as it occurs without \dss being aware or notified; data is collected as soon as \dss enter the range of the collecting device.
This is known as \emph{passive} data collection.
The GDPR imposes restrictions on (passive) data collection; namely it can only be carried out if \dss provide explicit consent.
We describe a refinement for passive data collection, which makes it possible for \dss to provide (or not) explicit consent.
\looseness -1

Consider a Bluetooth tracking system installed in a shopping mall.
The system is composed by data collection devices (\dc devices) that automatically record the Bluetooth MAC address of \ds devices within their range---\eg, smartphones, smartwatches, wireless headsets, etc.
Direct communication cannot be applied here.
\dc devices can \emph{only} collect data broadcast by \ds devices, and \dc devices cannot communicate the \dc policy prior to collecting the data.
Nevertheless, \ds devices can communicate data items and policies via
Bluetooth.
We exploit this feature to make it possible for the \dss to provide
consent.
\looseness -1

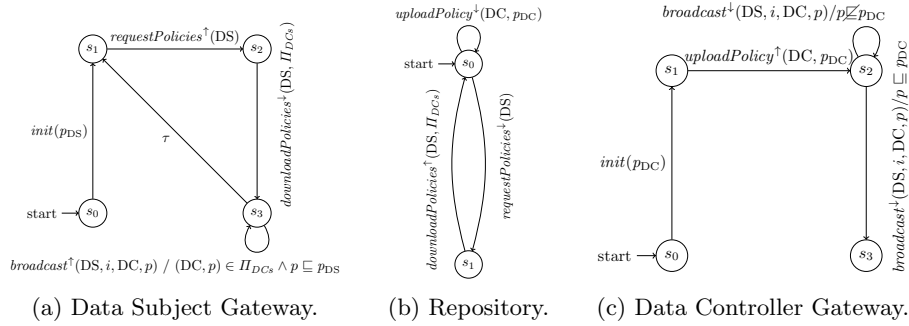
\begin{figure*}[t!]
    \begin{subfigure}[b]{0.38\textwidth}
    \centering
        \resizebox{\linewidth}{!}{
            \begin{tikzpicture}[every text node part/.style={align=center},on grid, auto]
        \node (s_0) [initial, initial where=left, circle, draw]%
                    {$s_0$};
        \node (s_1) [circle, draw]%
                    [above = 4cm of s_0]
                    {$s_1$};
        \node (s_2) [circle, draw]%
                    [right = 4cm of s_1]
                    {$s_2$};
        \node (s_3) [circle, draw]%
                    [below = 4cm of s_2]%
                    {$s_3$};
        \path[->] (s_0) edge node {$\mathit{\mathit{init}(\pds)}$} (s_1);
        \path[->] (s_1) edge node {$\sendAction{requestPolicies}(\ds)$} (s_2);
        \path[->] (s_2) edge node [rotate=90, xshift=-20mm, yshift=-7mm] {$\receiveAction{downloadPolicies}(\ds,\Pi_{\mathit{DCs}})$} (s_3);
        \path[->] (s_3) edge node {$\tau$} (s_1);
        \path[->] (s_3) edge [loop, min distance=10mm, in=240,out=300] node [xshift=-20mm]
                  {$\sendAction{broadcast}(\ds,i,\dc,p)~/~(\dc,p) \in \Pi_{\mathit{DCs}} \wedge p \subsumes \pds$} (s_3);
                \end{tikzpicture}
        }        
        \caption{Data Subject Gateway.}
        \label{fig:dsg-indirect}
    \end{subfigure}
    \hspace{3mm}
    \begin{subfigure}[b]{0.18\textwidth}
        \centering
        \resizebox{\linewidth}{!}{
            \begin{tikzpicture}[every text node part/.style={align=center},on grid, auto]
        \node (s_0) [initial, initial where=left, circle, draw]%
                    {$s_0$};
        \node (s_1) [circle, draw]%
                    [below = 5cm of s_0]
                    {$s_1$};
        \path[->] (s_0) edge [loop, min distance=10mm, in=60,out=120] node {$\receiveAction{uploadPolicy}(\dc,\pdc)$} (s_0);
        \path[->] [bend left = 15] (s_0) edge node [rotate=90, xshift=-15mm, yshift=-5mm] {$\receiveAction{requestPolicies}(\ds)$} (s_1);
        \path[->] [bend left = 15] (s_1) edge node [rotate=90, xshift=20mm, yshift=5mm] {$\sendAction{downloadPolicies}(\ds,\Pi_{\mathit{DCs}})$} (s_0);
\end{tikzpicture}
        }
        \caption{Repository.}
        \label{fig:repo-indirect}
    \end{subfigure}
    \hspace{3mm}
    \begin{subfigure}[b]{0.36\textwidth}
        \centering
        \resizebox{\linewidth}{!}{
            \begin{tikzpicture}[every text node part/.style={align=center},on grid, auto]
        \node (s_0) [initial, initial where=left, circle, draw]%
                    {$s_0$};
        \node (s_1) [circle, draw]%
                    [above = 4cm of s_0]
                    {$s_1$};
        \node (s_2) [circle, draw]%
                    [right = 4.2cm of s_1]
                    {$s_2$};
        \node (s_3) [circle, draw]%
                    [below = 4cm of s_2]%
                    {$s_3$};
        \path[->] (s_0) edge node {$\mathit{\mathit{init}(\pdc)}$} (s_1);
        \path[->] (s_1) edge node {$\sendAction{uploadPolicy}(\dc,\pdc)$} (s_2);
        \path[->] (s_2) edge [loop, min distance=10mm, in=60,out=120] node [xshift=-20mm] {$\receiveAction{broadcast}(\ds,i,\dc,p)/p \cancel{\subsumes} \pdc$} (s_2);
        \path[->] (s_2) edge node [rotate=90, xshift=-25mm, yshift=-7mm] {$\receiveAction{broadcast}(\ds,i,\dc,p) / p \subsumes \pdc$} (s_3);
\end{tikzpicture}
        }
        \caption{Data Controller Gateway.}
        \label{fig:dcg-indirect}
    \end{subfigure}
\caption{Indirect Communication Program Graphs.} 
\label{fig:indirect_pgs}
\end{figure*}

\textit{Policy Repository}.
\dcs can upload their policies to a \emph{public repository}.
These repositories store and send policies, but do not modify them.
The repository could be hosted by the \dc or governmental entities such as Data Protection Agencies.
\ds devices download \dc policies from repositories \emph{before} entering the operation range of the \dc device.
Otherwise, \dcs\ cannot store the \ds data (see details below).
Fig.~\ref{fig:repo-indirect} shows the policy repository PG.
The variable \pdc\ corresponds to $\policies(\dc,\dc)$.
The database of all received policies is denoted as $\Pi_{\dcs}$.
This is a temporary variable to support indirect exchange of policies between \dcs and \dss; not part of the abstract semantics.
The PG includes the current state $\sigma$ (not in abstract semantics).
The \Effect\ function is defined by synchronization actions; except for the action $\receiveAction{uploadPolicy}(\dc,p_\dc)$ that additionally updates the repository database with $p_\dc$,
$\Pi_\dcs \cup \{(\dc,p_\dc)\}$.
The repository can either receive \dc policies, $\receiveAction{uploadPolicy}(\dc,p_\dc)$, or send policies to \dss, $\sendAction{downloadPolicies}(\ds,$ $\Pi_{\mathit{DCs}})$, upon receiving a request $\receiveAction{requestPolicies}(\ds)$.
\looseness -1

\textit{Data Subject Gateway (\ds program)}.
This program broadcasts a data item and \ds policy together.
Fig.~\ref{fig:dsg-indirect} shows the PG.
The \ds policy $\pds$ corresponds to $\policies(\ds,\ds)$.
$\Pi_\dcs$ represents the repository database, and $\sigma$ the
PG current state (none in the abstract semantics).
\Effect is defined as 
$\Effect(init(\pds,i), \eta) = \vareval[\pds \mapsto {\texttt{policy\_init()}}, i \mapsto {\texttt{data\_init()}}]$, 
and 
$\Effect(\tau,\eta) = \eta$.
\looseness -1

The PG starts with an initialization action, $\mathit{init}(\pds)$,
where the \ds defines her policy and data item $i$ is recorded.
The function \texttt{policy\_init()} retrieves the policy from the \ds, and, \texttt{data\_init()} initializes $i$ to the Bluetooth MAC address of the device.
Then, it requests from the policy repository the database of \dc policies, $\sendAction{requestPolicies}(\ds)$.
Upon receiving them, $\receiveAction{downloadPolicies}(\ds,\Pi_{\mathit{DCs}})$, the data subject gateway starts to broadcast the data item $i$, receiver \dc, and policy
$p$---this might be implemented as in~\cite{trustcom:2019}.
This step iterates over all policies in $\Pi_\dcs$ that comply with the \pds.
Non-deterministically (or, alternatively, triggered by the \ds), the
data subject gateway can update its policy database---action $\tau$ in
the PG.
\looseness -1

\textit{Data Controller Gateway (\dc program)}.
This program can upload the \dc policy to the policy repository, and collect and process \ds data according to their policy.
We omit transfers as they are identical to that in Fig.~\ref{fig:dcg-direct}; it simply requires adding $s_4$ and the connecting edges.
Fig.~\ref{fig:dcg-indirect} shows the PG.
Variables $p$ and $i$ correspond to the policy and data item sent by the \ds: $ (i,\ds,p) \in \receiveddata(\dc)$ in the abstract semantics.
As in previous PGs, there is a current state variable $\sigma$ (not in the abstract semantics).
The $\Effect$ function is defined by synchronization actions and a policy initialization machine-level function.
\looseness -1

The PG starts by initializing the \dc policy, \pdc.
In order for \dss to access \pdc, the PG uploads it to the policy repository, $\sendAction{uploadPolicy}(\dc,\pdc)$.
Then, the data controller gateway starts collecting MAC addresses and corresponding policies from \ds devices within its range of operation.
When the received policy does not conform \pdc (loop in $s_2$), the data item and policy are discarded, and the PG continues collecting data.
Note that the \dc device may receive policies for other \dcs, in such a case the policy will fail to subsume that of the \dc.
We remark that, in our setting, policies are not secret, \eg, there is no problem in Amazon learning the policy that Alice defined for Google.
\looseness -1


\section{Verification of Legal Privacy Requirements}
\label{sec:verification}
We use model-checking to verify that:
\begin{inparaenum}[i)]
\item the abstract semantics complies with privacy requirements inspired by the GDPR; and
\item the PGs in \S\ref{sec:refinement} refine the abstract semantics---thus, satisfying the same privacy requirements. 
\end{inparaenum}
We mechanize the abstract semantics and PGs in \tlaplus~\cite{TLA+}, and use the \tlaplus\ Toolbox~\cite{TLA+toolbox} to automatically verify privacy requirements and refinements.
The \tlaplus\ code is available in~\cite{pilot-tla-doi,formalization-repo}.
\looseness -1

\textit{\tlaplus\ Mechanization}.
The abstract semantics is directly encoded as defined in Table~\ref{tab:os-com-events}---\tlaplus\ includes set theoretic operations as well as the next state operator \emph{prime} ($\prime$).
As for PGs, the only challenge is encoding synchronization actions (see Def.~\ref{def:parallelcomposition}).
To this end, we use a variable, \msgs, modeling the set of messages in transit.
Actions $\sendAction{a}(\vec{x})$ and $\receiveAction{a}(\vec{x})$ add and remove an element $\langle a , \vec{x} \rangle$ in \msgs, respectively.
Note that this type of asynchronous communication differs from the synchronous composition in Def.~\ref{def:parallelcomposition}. 
Consequently, the PGs in \S\ref{sec:refinement} are designed so that, if the PG is in a state waiting receive data (i.e., a state with an outgoing transition labeled with a receiving action), then the PG cannot make progress unless the corresponding message is sent. 
Hence, they exhibit the same synchronization behavior as specified in Def.~\ref{def:parallelcomposition}.

This formalization serves as a guideline for software engineers to verify privacy requirements and refinements.
\tlaplus\ has been used in real case studies by engineers in large companies such as Amazon~\cite{DBLP:journals/cacm/NewcombeRZMBD15}.
We demonstrate the use of the model-checker in the \tlaplus\ Toolbox to verify privacy requirements and refinements on a model comprising: a \ds, a data item (owned by the \ds), two \dcs, three \pilot\ policies $p_1$, $p_2$, $p_3$ (randomly selected by devices in their initialization steps) with $p_1 \subsumes p_2$.
As before, these policies are always active. This corresponds to worse-case analysis (as required in security and privacy applications) as it permits data communication for all data collection and transfer rules.
We chose this model as it is the smallest model triggering all events (which are explored non-deterministically by the model-checker).
To improve the performance of model-checking, we abstracted the syntax of \pilot\ policies.
The relation $\subsumes$ is explicitly encoded as a partial order in the model.
Model-checking privacy requirements and refinements on these models takes less than 10 seconds on a 1x1.7GHz virtual machine with 4G of RAM.
Unfortunately, verifying models including more entities, data items or policies introduces a scalability problem due to state explosion.
However, we conjecture that these additions will not change the results.
Note that: 
\begin{inparaenum}[i)]
  \item consent properties concern each data item separately (i.e., the handling of consent for one item does not affect another);
  \item we consider (non-deterministically) independent and dependent (in terms of $\subsumes$) policies for each item; and 
  \item all communication events are (non-deterministically) triggered.
\end{inparaenum}
Our future work includes mechanizing a formal proof of this conjecture.
\looseness -1

\subsubsection{Legal Privacy Requirements.}
\label{subsec:model-checking-properties}

We verify that: 
\begin{inparaenum}[i)]
\item the abstract semantics do not allow \dcs\ to violate
\ds\ policies; and
\item one of the necessary conditions for informed consent according to the GDPR.
\end{inparaenum}
\begin{privacyrequirement}[\ds Policy Compliance]\it
  \dcs\ always follow a policy $p_i$ for processing a data item $i$ that complies with the policy defined by the \ds, $p_\ds \; \subsumes \; p_i$.
\end{privacyrequirement}
This requirement ensures that \dcs\ cannot bypass the
constraints imposed by \dss\ in processing their data.
Formally, given device $d$, data controller $dc$, and data subject
$ds$,
$$
\text{if } (d,i,p_i) \in \receiveddata_{dc} \text{ and } \pds = \policies(ds,ds)  \text{ then } p_i \subsumes \pds
$$
with $\type(i) \podata \pds.t$.
The requirement is formalized and successfully verified~\cite{pilot-tla-doi}.
This requirement is of utmost importance as it ensures that \dss\ choices are never violated, and confirms the correctness
of the design of our semantics.
\looseness -1

The GDPR states that consent must be \emph{informed}.
Intuitively, this requirement means that \emph{before} data is
collected the \ds\ must be informed about how data will processed and
whether it will be transferred.
This requirement is of special interest, as there exist
implementations of cookie banners that failed to comply with informed
consent~\cite{cookiebannerproblems}.
Below we state two necessary privacy requirements to enforce
informed consent.
\looseness -1

\begin{privacyrequirement}[Informed Consent for Data Collection]
  {\it Be\-fo\-re a \dc\ receives data item $i$, the owner of $i$ must have
  received a \pilot\ policy from the \dc.}
\end{privacyrequirement}
Formally, given a \ds\ \sndr, \dc\ \rcv, and data item $i$,
\begin{align*}
\text{if } (\sndr,i,\_) \in \ReceivedDataSet{\rcv} \text{ and } \owner(i) = \sndr &\\ 
&\hspace*{-40mm} \text{ then } (\rcv,p) \in \PolicySet{\owner(i)} \text{ and } \type(i) \podata p.\dt.
\end{align*}
This property states that a \ds has provided
informed consent for data collection iff every time that a \dc
collects data from a \ds, the \ds has received a policy for
the collected item from the \dc.
This privacy requirement concerning informed consent is successfully
verified (see \cite{pilot-tla-doi}).
\looseness -1

We have verified that \pilot abstract semantics enforces two privacy properties which are required by the GDPR.
This result shows that software engineers (such as cookie banner developers) could use \pilot\ abstract semantics as a high-level design for their implementations.
Furthermore, as we describe below, they can use our formalization to
verify that their implementations satisfy our abstract semantics; thus
satisfying the privacy requirements discussed here.
\looseness -1

\subsubsection{Refinements}
\label{subsec:model-checking-refinements}

In \tlaplus, refinement (or implementation) amounts to logical
implication; the so called \emph{refinement mappings}~\cite{refmappings91}.
Formally, a specification $I$ implements an abstract specification
$S$ iff $I \Longrightarrow S$.
This proof technique requires mapping state variables in the
specification to state variables in the implementation.
We described this mapping for all the PGs in \S\ref{sec:refinement}---see our \tlaplus mechanization for details~\cite{pilot-tla-doi}.
\looseness -1

Let $\abstractsemantics$ denote the abstract semantics specification
(Table~\ref{tab:os-com-events}), $\directrefinement$ the specification
of direct communication (PGs in
Fig.~\ref{fig:direct_pgs}), and
$\indirectrefinement$ the specification of indirect communication (PGs
in Fig.~\ref{fig:indirect_pgs}).
We successfully verify that
\begin{inparaenum}[i)]
\item $\directrefinement \hspace*{-1mm} \Longrightarrow \hspace*{-1mm} \abstractsemantics$; and 
\item $\indirectrefinement \hspace*{-1mm} \Longrightarrow \hspace*{-1mm} \abstractsemantics$ (see~\cite{pilot-tla-doi}).
\end{inparaenum}
\looseness -1

The most important consequence of this result is that both refinements are guaranteed (proven) to enforce the privacy requirements above.
Furthermore, new privacy requirements proven on the abstract semantics will also hold in the refinements.
This is useful in practice as formal reasoning on the abstract semantics is much easier; the abstract semantics omit implementation details unnecessary for reasoning about privacy.
\looseness -1


\section{Related work and concluding remarks}
\subsubsection{Related work}

Several languages have been proposed to formally define privacy policies  \eg, S4P, CI, PrivacyAPI, PrivacyLFP and \pilot~\cite{S4Pbmb10,BDMNpcifa06,MGLpaactavlpp06,DGJKDelshgpl10,PardoLeMetayerDBSec19}.
Generally, these works focus on reasoning about system behavior with respect to privacy policies and regulations.  
However, none of them address the gap between formal definition and implementation.
Our refinement method (\S\ref{sec:refinement}) contributes to bridging this gap, and demonstrates it for the \pilot language---although our method can be applied to any privacy policy language whose policies and system behavior can be formally defined.
In~\cite{PardoLeMetayerDBSec19}, the authors used the SPIN model-checker~\cite{spin} to answer privacy risk queries.
However, the model used for verification was based on an informal description of system events.
Here we formalize the abstract semantics in \tlaplus, we formalize two GDPR requirements for the implementation of consent, and introduce a method to specify and verify refinements. 
\looseness -1

Different enforcement implementations for the P3P privacy policy language~\cite{FCESMCpdow08} have been studied via refinements~\cite{DBLP:journals/scp/PapanikolaouCG12}.
Papanikolaou et al. translate P3P policies into CSP~\cite{cspHoare78} and refinement is interpreted as a comparison between policies~\cite{DBLP:journals/scp/PapanikolaouCG12}.
Instead, we use a system level notion of refinement.
We verify that  the behavior of the implementation is equivalent to that of an abstract semantics satisfying GDPR requirements for informed consent.
\looseness -1

Some programming languages are designed to embed the enforcement of privacy requirements~\cite{TokasO20,DBLP:conf/csfw/KaramiBJ22}.
In~\cite{TokasO20}, the authors propose a programming language that includes consent management.
However, it does not include notions of data transfer and \dc policy. Therefore, it cannot be used to address the GDPR requirements in this paper.
DPL~\cite{DBLP:conf/csfw/KaramiBJ22} covers more GDPR requirements than our work, but its operational semantics is higher level than PGs.
The system configuration operates on objects but it does not specify how different devices execute each program---as we did for our PGs.
Our refinement method could be used to refine DPL semantics to lower level implementations.
\looseness -1

Hublet et al. introduced an enforcement of GDPR requirements for web applications~\cite{DBLP:conf/esorics/HubletBK23}.
The authors present an enforceable GDPR specification for web applications written in Metric First-Order Temporal logic.
Their work covers more GDPR requirements than the work presented here, but it is limited to web applications.
Our method does not impose constraints on architecture.
For instance, we showed a refinement of a Bluetooth-based communication system in \S\ref{sec:refinement}.
\looseness -1

\subsubsection{Conclusion}
We proposed a method to implement and verify legal requirements for consent management.
The method aims to assist software engineers in designing and verifying systems that require consent management.
The design/implementation language is program graphs---extended state machines that are common in the education of software engineers.
We use a verification toolbox~\cite{TLA+toolbox} that has been used by software engineers in industry~\cite{DBLP:journals/cacm/NewcombeRZMBD15,DBLP:conf/icse/HackettRK23}.
We provided a formal abstract semantics for the \pilot\ language, and verified that it satisfies GDPR requirements for informed consent.
We introduced a notion of refinement as operational models to design the implementation of the devices in the system.
We demonstrated the use of refinements to implement direct and indirect communication systems.
We verified that the implementations refine \pilot\ abstract semantics, \ie, they ensure the same GDPR requirements for informed consent.
We mechanized the abstract semantics, refinements and privacy requirements in \tlaplus, and used model-checking for verification.
\looseness -1


\bibliographystyle{splncs04}
\bibliography{references}

\appendix
\section{Policy subsumption}
\label{sec:policy-subsumption}
We recall the definition of \emph{\pilot\ policy subsumption}~\cite{PardoLeMetayerDBSec19}.

\begin{definition}[Data Usage Rule Subsumption]
  Given two data usage rules
  $\dur_1 = \langle P_1, \rt_1 \rangle$ and
  $\dur_2 = \langle P_2, \rt_2 \rangle$,
  we say that $\dur_1$ \emph{subsumes} $\dur_2$, denoted as
  $\dur_1 \dursubsumes \dur_2$, iff
  \begin{inparaenum}[i)]
    \item $\forall p_1 \in P_1 \cdot \exists p_2 \in P_2 \text{ such that } p_1 \popurposes p_2$; and
    \item $\rt_1 \leq \rt_2$.
  \end{inparaenum}
  \label{def:dur-subsumption}
\end{definition}

\begin{definition}[Data Communication Rule Subsumption]
  Given two data communication rules
  $\dcr_1 = \langle c_1, e_1, \dur_1 \rangle$ and
  $\dcr_2 = \langle c_2, e_2, \dur_2 \rangle$,
  we say that $\dcr_1$ \emph{subsumes} $\dcr_2$, denoted as
  $\dcr_1 \dcrsubsumes \dcr_2$, iff
  \begin{inparaenum}[i)]
    %
    %
    \item $e_1 \poentities e_2$; and
    \item $\dur_1 \dursubsumes \dur_2$.
  \end{inparaenum}
  \label{def:dcr-subsumption}
\end{definition}

\begin{definition}[\pilot Privacy Policy Subsumption]
  Given two \pilot privacy policies
  $\pi_1 = \langle t_1, \dcr_1, \TR_1 \rangle$
  and
  $\pi_2 = \langle t_2, \dcr_2, \TR_2 \rangle$,
  we say that $\pi_1$ \emph{subsumes} $\pi_2$, denoted as
  $\pi_1 \subsumes \pi_2$ iff
  \begin{inparaenum}[i)]
    \item $t_1 \podata t_2$;
    \item $\dcr_1 \dcrsubsumes \dcr_2$; and
    \item $\forall \tr_1 \in \TR_1 \cdot \exists \tr_2 \in \TR_2
    \text{ such that } \tr_1 \dcrsubsumes \tr_2$.
  \end{inparaenum}
  \label{def:specification-policy-subsumption}
\end{definition}

\section{Condition evaluation, Active Policies and Transfer rules}\label{sec:active-policy-transfer}
We recall the definitions for \emph{condition evaluation}, \emph{active policy}, and \emph{active transfer}~\cite{PardoLeMetayerDBSec19}.
%


\begin{definition}[Condition Evaluation]
Given the valuation function $\dataval$, device $d \in \Devices$, and condition $\phi \in \wfc$, \texttt{eval()} is defined as shown in Table~\ref{tab:eval}.
\begin{table}[h!]
  \begin{tabular}{rcl}
    $\eval{\dataval}{d}{\true}$ & $=$ &  $\texttt{true}$\\
    $\eval{\dataval}{d}{\false}$ & $=$ & $\texttt{false}$\\
    $\eval{\dataval}{d}{i}$ & $=$ & $\dataval(d,i)$\\
    $\eval{\dataval}{d}{c}$ & $=$ & $\hat{c}$\\
    $\eval{\dataval}{d}{f(t_1,t_2,\ldots)}$ & $=$ & $\hat{f}(\eval{\dataval}{d}{t_1},\eval{\dataval}{d}{t_2},\ldots)$\\
    $\eval{\dataval}{d}{t_1 \ast t_2}$ & $=$ &
    $\begin{cases}
      \eval{\dataval}{d}{t_1} ~ \hat{\ast} ~ \eval{\dataval}{d}{t_2} \\
      \hspace{20mm} \text{ if } \eval{\dataval}{d}{t_i} \not = \bot \\
      \bot \text{ otherwise}
    \end{cases}$\\
    $\eval{\dataval}{d}{\phi_1 \wedge \phi_2}$ & $=$ &
    $\begin{cases}
      \eval{\dataval}{d}{\phi_1} \text{ and } \eval{\dataval}{d}{\phi_2} \\ 
      \hspace{20mm} \text{ if } \eval{\dataval}{d}{\phi_i} \not = \bot \\
      \bot \text{ otherwise}
    \end{cases}$\\
    $\eval{\dataval}{d}{\neg \phi}$ & $=$ &
    $\begin{cases}
      \text{not } \eval{\dataval}{d}{\phi} \\
      \hspace{20mm} \text{ if } \eval{\dataval}{d}{\phi} \not = \bot \\
      \bot \text{ otherwise}
    \end{cases}$\\
  \end{tabular}
  \caption{Definition of \eval{\dataval}{d}{\phi}
  We use $\hat{c}$, $\hat{f}$ and $\hat{\ast}$ to denote the interpretation of constants, functions and binary predicates, respectively.
  We assume that these interpretations are the same across all devices.
  }
  \label{tab:eval}
\end{table}
\end{definition}


\begin{definition}[Active \pilot policy]
Given, devices $\sndr,\rcv \in \Devices$, policy $p \in \wfpp{}$, data item $i \in \DataItems$ and state $\mathit{st} \in \States$,  
$$
\begin{array}{c}
  \activePolicy(p, \send(\sndr,\rcv,i), \st) = \\
          \type(i) \podata \datatype \wedge
          \eval{\dataval}{\sndr}{\phi} \wedge
          \timestamp(\st, \send(\sndr,\rcv,i)) < \rt ~ \wedge \\
          \devtoent(rcv) \poentities e
\end{array}
$$
\noindent
where $p = (\datatype,\langle \phi, e, \langle \_, \rt \rangle \rangle,\_)$
and $\st = \langle \dataval, \_, \_ \rangle$.\qed
\end{definition}

\begin{definition}[Active transfer rule]
Given, devices $\sndr,\rcv \in \Devices$, policy $p \in \wfpp{}$, transfer rule $\tr \in p.\TR$, data item $i \in \DataItems$ and state $\mathit{st} \in \States$,  
$$
\begin{array}{c}
  \activeTransfer(\tr,
                  p,
                  \transfer(\sndr,\rcv,i),
                  \st) = \\
  \timestamp(st,\transfer(\sndr,\rcv,i)) < \rt_{p} ~ \wedge
  \\
  \eval{\dataval}{\sndr}{\phi_{\tr}} \wedge
  \timestamp(\st,\transfer(\sndr,\rcv,i)) < \rt_{\tr} ~ \wedge
  \\
  \devtoent(\rcv) \poentities e_{\tr}
\end{array}
$$
\noindent
where
$\tr = \langle \phi_{\tr}, e_{\tr}, \langle \_, \rt_{\tr} \rangle \rangle$,
$p   = (\_, \langle \_, \_, \langle \_, \rt_{p} \rangle \rangle, \_ )$
and
$\st = \langle \dataval, \_, \_ \rangle$.\qed
\end{definition}


\end{document}